\documentclass[aps,amsfonts]{revtex4}
\usepackage{epsfig}
\usepackage{graphicx}
\usepackage{amsmath}
\usepackage{amsbsy}
\begin{document}
\newcommand{\ee}{\end{equation}}
\newcommand{\bb}{\begin{equation}}
\newcommand{\eqb}{\begin{eqnarray}}
\newcommand{\eqf}{\end{eqnarray}}
\def\x{\mathbf{x}}
\def\p{\mathbf{p}}
\def\ho{{\mbox{\tiny{HO}}}}
\def\sc{\scriptscriptstyle}
\title{$SU(2)$ Kinetic Mixing  Terms and Spontaneous Symmetry Breaking}
\author{J. L\'opez-Sarri\'on}
\email{justinux75@gmail.com}
\affiliation{Departamento de F\'{\i}sica, Universidad de Santiago de Chile,
Casilla 307, Santiago, Chile }
\author{Paola Arias}
\email{paola.arias@gmail.com}
\affiliation{Departamento de F\'{\i}sica, Universidad de Santiago de Chile,
Casilla 307, Santiago, Chile }
\author{J. Gamboa}
\email{jgamboa55@gmail.com}
\affiliation{Departamento de F\'{\i}sica, Universidad de Santiago de Chile,
Casilla 307, Santiago, Chile }

\begin{abstract}
The non-abelian generalization of the Holdom model --{\it i.e.} a theory with two gauge fields coupled to the
 kinetic  mixing term $g~ \mbox{tr}\left(F_{\mu \nu} (A) F_{\mu \nu} (B) \right)$--  is considered. Contrarily to the abelian case,
 the group structure $G\times G$ is explicitly broken to $G$. For $SU(2)$ this fact implies that the residual
 gauge symmetry as well as the Lorentz symmetry is spontaneusly broken. We show that  this mechanism provides of
 masses for the involved particles. Also, the model presents instanton solutions with a redefined coupling constant.
\end{abstract}
\maketitle

\section{Introduction}

The search for new extensions of the Standard Model has stimulated
the interest of the so called hidden sector of particles, which
could interact very weakly with other known particles. In the
context of string theory, the hidden sector appears naturally,
predicting additional $U(1)$ factors \cite{dienes,taylor, abel,okun},
with interesting phenomenological consequences \cite{murayama}.

This last idea has been explored in the abelian sector in \cite{holdom}, by considering two gauge fields interacting via a
renormalizable gauge interaction by means of the following
lagrangean

\bb
 {\cal L}= {\cal L}_0 + {\cal L}_{\mbox{int}}, \label{1}
 \ee
 where
 \bb
 {\cal L}_0= \frac{1}{4} F^2_{\mu \nu} (A)+ \frac{1}{4} F^2_{\mu \nu} (B), \label{01}
\ee
and
\bb
{\cal L}_{\mbox{int}} = \frac{g}{2} F_{\mu \nu} (A) F_{\mu \nu} (B), \label{02}
\ee
where $g$ is a dimensionless coupling constant, $A_\mu$, $B_\mu$ are abelian gauge fields  and the strength tensor
$ F_{\mu \nu} $ is defined as usual, {\it i.e.} $ F_{\mu \nu} (A)= \partial_\mu A_\nu - \partial_\nu A_\mu$.

Equation (\ref{1}) is the most general Lagrangian containing two gauge fields invariant under $U(1)\times U^{'}(1)$. Following
Holdom \cite{holdom} and others \cite{ring}, (\ref{1}) can be diagonalized very easily by using the transformation
\bb
B_\mu^{'} = B_\mu + g A_\mu. \label{2}
\ee
and (\ref{1}) becomes
\bb
{\cal L}=  \frac{1}{4} (1-g^2) F^2_{\mu \nu} (A) + \frac{1}{4}F^2 _{\mu \nu}(B^{'}), \label{3}
\ee
where the electric charge now is redefined as
$$
{\tilde e}^2 = \frac{1}{1-g^2}.
$$

This last result  was reached by Holdom \cite{holdom} and recently this question has
been revived in \cite{ring,lang} motivated by different reasons. However,
to the best of our knowledge, there are no studies concerning
to possible kinetic mixing in the non-abelian sector.

Thus the goals of the present research are the following: firstly to generalize (\ref{1}) to the non-abelian case and, secondly,
to explore the physical meaning of the  kinetic mixing terms. We will show below that this simple extension contains
unsuspected properties such as,  a) spontaneous symmetry breaking, and hence, appearance of Higgs
bosons and massive gauge bosons, and b) vacuum instantons solutions.

The paper is organized as follows; in section II the free non-abelian theory containing two gauge fields is considered. In
section III the kinetic mixing terms and its physical implications are discussed. Section IV is devoted to the
spontaneous symmetry breaking phenomenon and generation of mass for gauge bosons which is present in the model.
Section V contains a discussion on instantons
and finally we conclude in section VI with some final remarks and outlook.
\section{\lq \lq Free" non-abelian gauge field theory}

In order to expose our results, let us start by considering a model with two \lq \lq free" gauge fields described by the
Lagrangian
\bb
{\cal L}_0 = \frac{1}{4} {\mbox  tr} \left(F^2_{\mu \nu}(A)\right)  + \frac{1}{4} {\mbox  tr} \left(F^2_{\mu \nu}(B)\right),
\label{4}
\ee
where $F_{\mu \nu}(A)$ and $F_{\mu \nu}(B)$ are the field strengths defined as usual
\[
F_{\mu \nu}(A) = \partial_\mu A_\nu - \partial_\nu A_\mu + e_A [A_\mu,A_\nu],~~~~~~~~~~
F_{\mu \nu}(B) = \partial_\mu B_\nu - \partial_\nu B_\mu + e_B [B_\mu,B_\nu],
\]
with $A \equiv A^a T^a$ and $B\equiv B^b S^b$, the gauge potential fields associated to two identical copies of $SU(2)$  and
where
$T^a$ and $S^a$ are the generators of $SU(2) \times SU(2)$ in a given
representation. Thus, $T$ and $S$ are the generators of the first and the second copy of $SU(2)$ respectively.

The full symmetry group of the theory is  $SU(2) \times SU(2)$, or in other words this system is
invariant under the gauge transformations,
\bb
A_\mu\longrightarrow U_A^{-1} A_\mu U_A + e_A^{-1} U^{-1}_A\partial_\mu U_A, \label{5}
\ee
and,
\bb
B_\mu\longrightarrow U_B^{-1} B_\mu U_B + e_B^{-1} U^{-1}_B\partial_\mu U_B,\label{6}
\ee
where $U_A$ and $U_B$ are elements of $SU(2) \times SU(2)$, which act on the first and the second copy of $SU(2)$
respectively, and they are in general different.

Hence, one could define a unified connection ${\cal A}$ on $SU(2)\times SU(2)$, with normalized coupling constant $e=1$,
transforming as usual, {\it i.e.}
\bb
{\cal A}_\mu\longrightarrow {\cal U}^\dag {\cal A}_\mu{\cal U } +  {\cal U}\partial_\mu{\cal U}, \label{7}
\ee
on the  whole symmetry group.

Now, if we identify,
\bb
{\cal A}_\mu \equiv \alpha A_\mu\otimes I_s + \beta I_t\otimes B_\mu, \label{8}
\ee
where $\alpha$ and $\beta$ are constants, and  $I_{t,s}$ stands for the identity element of the group  in the representations
$T$ and $S$ respectively. Then, using (\ref{5})-(\ref{8}) one can check that the following consistency condition
$$ \alpha = e_A,\hspace{1cm}\beta = e_B,$$
must be fulfilled.

In order to find a Lagrangean for  ${\cal A}$ it is convenient to rewrite the generators of the group
as follows
\bb
{\cal T}^a = T^a\otimes I,\hspace{1cm} {\cal S}^a = I\otimes S^a, \label{10}
\ee
and then one can check that
\eqb
&& \left[{\cal T}^a,{\cal T}^b\right] = i\epsilon^{abc}{\cal T}^c, \nonumber
\\
&& \left[{\cal S}^a,{\cal S}^b\right] = i\epsilon^{abc}{\cal S}^c, \label{alge}
\\
&& \left[{\cal T}^a,{\cal S}^b\right] = 0.\nonumber
\eqf

However by defining the combinations
\bb
{\cal J}^a={\cal T}^a+{\cal S}^a,\hspace{1cm}{\cal K}^a= {\cal T}^a-{\cal S}^a,
\ee
which have the algebra
\eqb
&&[{\cal J}^a,{\cal J}^b] = i\epsilon^{abc}{\cal J}^c,\nonumber\\
&&[{\cal J}^a,{\cal K}^b] = i\epsilon^{abc}{\cal K}^c,\nonumber\\
&&[{\cal K}^a,{\cal K}^b] = i\epsilon^{abc}{\cal J}^c, \label{so4}
\eqf
which is --except by a sign  in the last equation-- the Lorentz group algebra. More precisely, this is the algebra for the
rotations
in $\mathbb R^4$, {\it i.e.}, $SO(4)$. With this in mind, we can extend the inner space in such a way that instead of having
$a,b=1,2,3$ we will have $\alpha, \beta = 1,2,3,4$.

In this language, one can unify the two kind of generators as
\bb
{\cal M}^{\alpha\beta} = \left\{\begin{array}{l l l}
{\epsilon^{abc}\cal J}^c& {\rm if}& \alpha=a,\beta=b\\
{\cal K}^a&{\rm if}& \alpha=0,\beta=a
\end{array}\right \}
\ee
and, therefore, the algebra (\ref{so4}) is summarized as
$$[{\cal M}^{\alpha\beta},{\cal M}^{\gamma\delta}] = -i(\delta^{\alpha\gamma}{\cal M}^{\beta\delta} -\delta^{\beta\gamma}{\cal
M}^{\alpha\delta} +
\delta^{\beta\delta}{\cal M}^{\alpha\gamma} -
\delta^{\alpha\delta}{\cal M}^{\beta\gamma})$$

We know the representations of this group, which can be thought of as the Lorentz group representations, but changing
$\eta^{\mu\nu}$ for $-\delta^{\alpha\beta}$. In particular, the simplest nontrivial representation is given by the Dirac's matrices
in the four-dimensional Euclidean space, where we use $i\gamma^0$ for $\Gamma^0$ and the other matrices as usual, {\it
i.e.},
\bb \Gamma^\alpha=\left(\begin{array}{c c}
0& \sigma^\alpha\\
\bar\sigma^\alpha&0\end{array}\right)
\ee
where $\sigma^0=\bar\sigma^0=iI_2$, and $\sigma^a=-\bar\sigma^a$ are the Pauli matrices for $a=1,2,3$. Then, the
representation for the generators are,
\bb
{\cal M}^{\alpha\beta} = \frac{i}{4}[\Gamma^{\alpha},\Gamma^\beta].
\ee
In particular,
\bb
{\cal J}^a=\frac{1}{2}\left(\begin{array}{c c}
\sigma^a&0\\
0&\sigma^a\end{array}\right),\hspace{1cm}
{\cal K}^a=\frac{1}{2}\left(\begin{array}{c c}
\sigma^a&0\\
0&-\sigma^a\end{array}\right),
\ee
and the analogous of $\gamma^5$ becomes
\bb
\Gamma^5= -\Gamma^0\Gamma^1\Gamma^2\Gamma^3 = \left(\begin{array}{c c}
I_2&0\\
0&-I_2\end{array}\right).
\ee
This matrix has the property that $\{\Gamma^5,\Gamma^\alpha\}=0$, and ${\cal K}^a = \Gamma^5{\cal J}^a$. This means
that an element of $SU(2)\times SU(2)$ might be expressed like,
$$U=U_A\otimes U_B = {\rm exp}[i(\theta^a + \eta^a\Gamma^5 ){\cal J}^a], $$
for matrices $U_A$ and $U_B$ in $SU(2)$ of the form,
$$U_A = e^{i\xi_A^aT^a},\hspace{1cm}U_B = e^{i\xi_B^aS^a},$$
where the parameters $\theta^a= (\xi_A^a+\xi_B^a)/2$ and $\eta^a= (\xi_A^a-\xi_B^a)/2$, are written in terms of the
parameters $\xi^a_A$ and $\xi^a_B$ of the transformation $U_A$ and $U_B$ respectively.
Hence, we can write,
\begin{eqnarray}
F^a_{\mu\nu}({\cal A }) &=& \partial_\mu{\cal A}_\nu - \partial_\nu{\cal A}_\mu + [{\cal A}_\mu,{\cal A}_\nu]=e_A F^a_{\mu
\nu}(
A){\cal T}^a + e_B F^a_{\mu \nu}(B){\cal S}^a \nonumber\\
&=&\left(\frac{e_A F_{\mu\nu}^a(A)+e_B F_{\mu\nu}^a (B)}{2}\right) {\cal J}^a +
\left(\frac{e_A F_{\mu\nu}^a(A)-e_B F_{\mu\nu}^a (B)}{2}\right){\cal K}^a.
\nonumber
\end{eqnarray}
So, in this representation,
\bb
F_{\mu\nu}({\cal A})= \frac{1}2\left(\begin{array}{c c}
e_A F_{\mu\nu}^a(A)\sigma^a&0\\
0& e_B F_{\mu\nu}^a (B)\sigma^a\end{array}\right). \label{FA}
\ee

Then with these facts in mind, one can see that the Lagrangean  ${\cal L}_0$ can be written in terms of the curvature
associated to
${\cal A}$. It reads,
\eqb
{\cal L}_0 &=& a~ {{\mbox tr}}F^2_{\mu \nu}({\cal A}) + a_5 ~{{\mbox tr}} \left( F_{\mu \nu}({\cal A})\Gamma^5
F_{\mu \nu}({\cal A})\right), \label{calA}
\\
&=& (a + a_5)\alpha^2 F^2_{\mu \nu}(A) + (a-a_5)\beta^2 F^2_{\mu \nu}(B), \label{calA1}
\eqf
where the coefficients $a$ and $a_5$
are defined as
\eqb
a &=& \frac{1}{8}\left(\frac{1}{e_A^2}+\frac{1}{e_B^2}\right), \label{a}
\\
a_5 &=& \frac{1}{8}\left(\frac{1}{e_A^2}-\frac{1}{e_B^2}\right)  \label{aga}
\eqf

The Lagrangean (\ref{calA}) and (\ref{calA1}), of course, are equivalent to (\ref{4}) with that choice for $a$ and $a_5$.
At this level one should note the following; firstly one can describe the \lq \lq free" case using two formalisms, namely, in
terms of
${\cal A}$
or in term of the pair $(A,B)$ and both are  equivalent descriptions. Secondly, these two descriptions can be  physically
interpreted
following a simple formal analogy with fermions.

Indeed, let us assume a free massless fermion field which is described either by chiral fields or by Dirac ones.
In the first case the Lagrangean for the chiral fields is
\bb
{\cal L}= \alpha_L ~\psi^\dagger_L \left(\partial_0 + {\vec \sigma}.\nabla \right) \psi_L +
\alpha_R  ~\psi^\dagger_R ~\left(\partial_0 - {\vec \sigma}.\nabla \right) \psi_R, \label{chi}
\ee
which is explicitly invariant under the chiral symmetry $SU(2)_L\times SU(2)_R$.

The second  possibility is to use Dirac fields and the Lagrangean reads
\bb
{\cal L} =\frac{(\alpha_L+\alpha_R)}{2} {\bar \psi}{\partial \hspace{-.6em} \slash \hspace{.15em}} \psi +
\frac{(\alpha_L-\alpha_R)}{2} {\bar \psi}\gamma_5 {\partial \hspace{-.6em} \slash \hspace{.15em}} \psi,  \label{chi1}
\ee
which is invariant under the chiral  and gauge symmetry transformations
\eqb
\psi^{'} (x) &=& e^{i \alpha_5 \gamma_5} \psi (x), ~~~~~~~~~~~{\bar \psi}^{'} (x) =  {\bar \psi}^{'} (x)e^{i \alpha_5 (x)
\gamma_5},
\label{si1}
\\
\psi^{'} (x) &=& e^{i \alpha (x)} \psi (x), ~~~~~~~~~~~{\bar \psi}^{'} (x) =  {\bar \psi}^{'} (x)e^{-i \alpha (x)}, \label{si2}
\eqf
and, obviously, the Dirac version is also invariant under $SU(2) \times SU(2)$. So either versions are equivalent.

In our case at hands the fields $A$ and $B$ are the analogous of chiral fields $\psi_L$ and $\psi_R$ and --only in this sense,
of course--
(\ref{4}) is a chiral description for two gauge fields. The field ${\cal A}$ is the analogous of a Dirac field with
${\bar \psi}\gamma_5 {\partial \hspace{-.6em} \slash \hspace{.15em}} \psi$ playing the role of
${{\mbox tr}} \left( F({\cal A})\Gamma^5 F({\cal A})\right)$.

Thus, an interesting point is the following; if a mass term in a fermionic theory breaks chiral symmetry, then,  what is the
analogous of a mass term in a gauge field theory as is discussed here?, and what are the physical implications?.  We will
answer these questions in the next section.

\section{Including Non-abelian Kinetic Mixing Terms}

Following the analogies discussed above we will consider the analogous of a \lq \lq mass" term for a theory with two non-
abelian gauge fields. This term should break partially the gauge symmetry in the same sense that
$SU(2)\times SU(2) \sim SU(2)$ in the fermionic case when a term like
$m(\psi^\dagger_L \psi_R + \psi^\dagger_R \psi_L)$ or $m{\bar \psi}\psi$ is added to (\ref{chi}) or (\ref{chi1}).

In a field theory involving two gauge fields, however, this partial gauge symmetry breaking has important physical
consequences as we will see below.

It is not difficult to see that this mass term must be
\bb
{\cal L}_I = \frac{g}2 {\mbox tr}\left(F_{\mu\nu}(A)F_{\mu\nu}(B)\right). \label{inte}
\ee
which, in turn, can be thought of as a non-abelian generalization of the Holdom and Okun model outlined in the introduction.

This term breaks  partially the gauge symmetry $SU(2)\times SU(2)$  because if
we perform the transformations (\ref{5})-(\ref{6}), one finds
\bb
{\cal L}_I \longrightarrow \frac{g}2 F^a_{\mu \nu}(A)\Lambda^{ab}(U^{-1}_A)\Lambda^{bc}(U_B)F^c_{\mu \nu}(B),
\label{chang}
\ee
where $\Lambda$ is in the adjoint representation of the $SU(2)$.

Thus,  we must restrict ourselves to those
transformations in $SU(2)\times SU(2)$ such that
\bb
U_A = U_B,  \label{restric}
\ee
in order to keep them as symmetry transformations. Hence the residual  symmetry becomes equivalent to $SU(2)$.

By writting the ${\cal A}$ field as,
\bb
{\cal A_\mu} ={\cal A}^{\alpha\beta}{\cal M}^{\alpha\beta}= Z^a_\mu {\cal J}^a + W_\mu^a {\cal K}^a, \label{con1}
\ee
where we have defined,
\bb
Z^a_\mu\equiv\frac{1}2\epsilon^{abc}{\cal A}^{bc}\equiv e_A A^a_\mu +e_B B^a_\mu, \label{con2}
\ee
and,
\bb
W^a_\mu\equiv{\cal A}^{0a}\equiv e_A A^a_\mu - e_B B^a_\mu. \label{con3}
 \ee
one can see that under a gauge transformation that preserves the mass term, namely with $U_A=U_B\equiv U$, the fields
$Z$ and $W$ transform as follows,
$$Z_\mu \rightarrow U^{-1}Z_\mu U + U^{-1}\partial_\mu U$$
and,
$$W_\mu \rightarrow U^{-1}W_\mu U $$
This says that $Z$ is a gauge potential under the residual gauge symmetry, and $W$ transforms in the adjoint
representation.

In order to write down the mixing term in terms of the ${\cal A}$ field, and hence in terms of the $Z$ and $W$ fields, let us
consider the the product $\Gamma^0F_{\mu\nu}({\cal A})\Gamma^0$.  It is actually easy to see that,
$$\Gamma^0F_{\mu\nu}({\cal A})\Gamma^0 =
-\frac{1}2\left(\begin{array}{c c}
e_B F_{\mu\nu}^a(B)\sigma^a&0\\
0& e_A F_{\mu\nu}^a (A)\sigma^a\end{array}\right).
$$
And, then, the mixing term can be rewriten as,
$${\cal L}_I = \frac{g}2 F_{\mu\nu}^a(A)F^{a\mu\nu} =-\frac{g}{2} {\mbox tr} \left( F_{\mu \nu}({\cal A}) \Gamma^0 F^{\mu
\nu}({\cal A}) \Gamma^0\right).$$
It is worthy to notice that the special role of the $\Gamma^0$ matrix of these term mimics the role of the $\gamma^0$ matrix in
the fermionic analogy suggested above, and this fact justifies the  ''massive" name for this term.

The full Lagrangean of this \lq \lq massive" model in terms of the connection ${\cal A}$ is, then,
\bb
{\cal L} = \frac{a}{4} {\mbox tr} \left( F^2 _{\mu \nu}({\cal A}) \right) +
\frac{a_5}{4} \left( F_{\mu \nu}({\cal A}) \Gamma^5 F^{\mu \nu}({\cal A})\right)
-\frac{g}{2} {\mbox tr} \left( F_{\mu \nu}({\cal A}) \Gamma^0 F^{\mu \nu}({\cal A}) \Gamma^0\right).  \label{u1}
\ee
Following the fermionic analogy, $SU_5(2)$  invariance is
broken in the same sense as chiral symmetry is broken in a massive fermionic field theory.

As we will see in the next section, some physical consequences of the model described by (\ref{u1}) can be understood more
easily  by expressing (\ref{u1}) in terms of $Z^a_\mu$ and
$W^a_\mu$. Then in terms of these fields the strength tensor is

\bb
F_{\mu\nu}({\cal A})=\left(F^a_{\mu\nu}(Z) + [W_\mu,W_\nu]^a\right){\cal J}^a + \left( (D_\mu W_\nu)^a- (D_\nu
W_\mu)^a\right) {\cal K}^a \label{FZW}
\ee
\bb
\Gamma^5 F_{\mu\nu}({\cal A})=\left( (D_\mu W_\nu)^a- (D_\nu W_\mu)^a\right){\cal J}^a +  \left(F^a_{\mu\nu}(Z) + [W_
\mu,
W_\nu]^a\right){\cal K}^a, \label{FZW1}
\ee
and,
\bb
-\Gamma^0 F_{\mu\nu}({\cal A})\Gamma^0=
\left(F^a_{\mu\nu}(Z) + [W_\mu,W_\nu]^a\right){\cal J}^a - \left( (D_\mu W_\nu)^a- (D_\nu
W_\mu)^a\right) {\cal K}^a \label{FZW2}
\ee
where the covariant derivative is $D_\mu W^a_\nu = \partial_\mu W^a_\nu + [ Z_\mu,W_\nu]$.
Thus,
\eqb
{\cal L} &=& \frac{1}{8e_A^2}\left\{ \left( F^a_{\mu\nu}(Z) + [W_\mu,W_\nu]^a\right) + \left( (D_\mu W_\nu) - (D_\nu
W_\mu)\right)\right\}^2  \nonumber
\\
&+&\frac{1}{8e_B^2}\left\{ \left( F^a_{\mu\nu}(Z) + [W_\mu,W_\nu]^a\right) - \left( (D_\mu W_\nu) - (D_\nu
W_\mu)\right)\right\}^2 \nonumber
\\
&+& \frac{g}2 \left(F^a_{\mu\nu}(Z) +  [W_\mu,W_\nu]^a
\right)^2 - \frac{g}2\left( (D_\mu W_\nu)^a- (D_\nu W_\mu)^a\right)^2, \label{fina}
\eqf
describes the full dynamics of a gauge field theory including the \lq \lq massive" term (\ref{inte}).
\section{Spontaneous Symmetry Breaking and Mass for $Z^a_\mu$}
The Lagrangean (\ref{fina}) contains several interesting physical properties as we will see in this section. The fact that we
explicitly broke the full symmetry has  a non trivial  consequence; the residual $SU(2)$ symmetry is spontaneusly broken. As
we pointed out at the end of the last section,  $Z^a_\mu$ is a genuine gauge potential whereas $W^a_\mu$ is a vector field
playing a role similar to the scalar one in the Higgs model.

To prove this statement, let us consider the lagrangean (\ref{fina}) with $Z$ put to zero. Then, the potential energy for $W$,
neglecting the spatial derivatives of $W$, is given, in the Euclidean space,  by,
\begin{equation}
V[W]=\int d^4x\, {\cal V}(W)=\int d^4x\,{\cal L}(Z,W)\vert_{Z=\partial W=0}, \label{pot1}
\end{equation}
with ${\cal V}$
\begin{equation}
{\cal V}(W) =[W_\mu,W_\nu]^a[W^\mu,W^\nu]^a = \gamma \left[(\vec W^c\cdot \vec W^c)^2 - (\vec W^b\cdot \vec W^c)^2
\right]
\end{equation}
where the notation ${\vec W}^c\cdot {\vec W}^c$ means ${W^c}_\mu{W^c}_\mu$ and so on. Also, we have defined the
constant
$$\gamma \equiv \left(\frac{1}{8e^2_A} + \frac{1}{8e_B^2} + \frac{g}2\right).$$

Now, we claim a nonvanishing vacuum expectation value for the field $W$, and then we redefine it in order to have
expectation values on the vacuum equal to zero for the physical fields $\omega$, {\it i.e.},
\begin{equation}
W^a_\mu = v^a_\mu + \omega^a_\mu
\label{lifting}
\end{equation}
To see the consistency of the above statement,
we
must see if it corresponds to an extremal point for the potential, {\it i.e.}, we must impose the condition,
\begin{equation}
\left.\frac{\partial {\cal V}}{\partial W^a_\mu}\right\vert_{W=v_0} = 0.
\end{equation}

This expression produces the set of equations,
\begin{equation}
(\vec v^b\cdot \vec v^b) \vec v^a - (\vec v^a\cdot\vec v^b)\vec v^b = 0,\hspace{1cm} a=1,2,3.
\end{equation}

It is easy to see --although not quite straightforward-- that these equations have the general solution,
\begin{equation}
(v_0)^a_\mu = v \hat\lambda^a \hat e_\mu,
\label{solu}
\end{equation}
where $v$ is an undetermined constant, $\hat\lambda^a$ and $\hat e_\mu$ are the components of unitary vectors in the
inner and Euclidean spaces respectively, {\it i.e.},
$$ \sum_{a=1}^3(\hat \lambda^a)^2 = 1 = \sum_{\mu=1}^4 (\hat e_\mu)^2. $$
So far, we have supposed that we are working in the Euclidean space, but, in short, we will see how it works for the
Minkowski space.

The next question is whether or not these vacua are stable. In order to answer  this question one must check the sign of the
mass matrix, namely, to see that,
\begin{equation}
\frac{1}2\left.\frac{\partial^2{\cal V}}{\partial W^a_\mu\partial W^b_\nu}\right\vert_{W=v_0}\omega^a_\mu \omega^b_\nu \geq
0. \label{massma}
\end{equation}
for any direction of $\omega$.
For our potential and the solutions (\ref{solu}), this yields,
\begin{equation}
\frac{1}2\left.\frac{\partial^2{\cal V}}{\partial W^a_\mu\partial W^b_\nu}\right\vert_{W=v_0} = 2\gamma v^2 (\delta^{ab} - \hat
\lambda^a
\hat\lambda^b)(\delta_{\mu\nu} - \hat e_\mu\hat e_\nu).
\end{equation}
However the condition (\ref{massma}) is satisfied only if,
\begin{equation}
g\geq -\left(\frac{1}{4e^2_A}+\frac{1}{4e^2_B}\right). \label{la1}
\end{equation}
At this point it is worth noting that in the Minkowski space the only well defined matrix, {\it i.e.} positive or negative, is when
$\hat e$ is a temporal-like vector, but because $\hat e^2=-1$ this matrix is negative defined. However, in the Minkowski space
the potential energy has a different sign from the Euclidean potential energy, and hence, the Minkowski energy potential is
positive defined for a  temporal-like vector $\hat e$, and the answer does not change for the Minkowski  space.

Thus assuming that (\ref{la1}) is hold, we obtain a nonvanishing mass for the $W$ physical fields, {\it i.e.} we obtain,
\begin{equation}
 m^2_W = 2v^2 \left(\frac{1}{4e^2_A}+\frac{1}{4e^2_B}+ g\right), \label{Mw}
\end{equation}
for six of the twelve degrees of freedom associated to $W^a_\mu$, and the other six massless excitations are Goldstone
bosons. This last fact can be seen more clearly by choosing an orthonormal basis such that, $\vec e_4 = \vec e$ and
$\vec e_i$ with $i=1,2,3$, a set of three orthonormal vectors to $\vec e$ and to each other. Also, by choosing
$\hat \lambda_3=\hat\lambda$ and
$\hat \lambda_A$ with $A=1,2$.
In this basis, the mass matrix is diagonal and it is written as,
$$\frac{1}2 m^2_W \delta_{AB}\delta_{ij}, $$
which says that only those components of $W$ of the form,
$$W_{\rm massive} = W^A_i \hat\lambda^A\otimes \hat e_i,$$
are massive.

Therefore although the mass of $W^a_\mu$ is hidden in the Lagrangean, the spontaneous symmetry breaking make it
explicit.

Next step is to consider the coupling to the $Z$ bosons. Following the standard arguments we can use the gauge freedom in
order to remove two massless $W$ bosons by choosing $U$ such that
$$ W^a_4 = \vec W^a\cdot \hat e = U^{-1} (\hat\lambda^3 + \omega){\cal J}^3U$$
where $U$ can be expressed as,
$$U=e^{i(\Psi^1{\cal J}^1 + \Psi^2{\cal J}^2)}$$
for some suitable functions $\Psi^1(x)$ and $\Psi^2(x)$.

With this transformation the $Z$ field changes as usual,
\begin{equation}
Z^a_\mu = U^{-1} \left[Z^{'}_\mu + \partial_\mu\right]U.
\end{equation}

The massless components $W^1_4$ and $W^2_4$ can be gauged out and, therefore, the mass matrix for this sector is,
$${\cal L}_{mass}(Z',W')=\frac{1}2 m^2_Z Z'^A_iZ'^B_j \delta^{AB}\delta_{ij},$$
where
\bb
m^2_Z = 2v^2 \left( \frac{1}{4e^2_A}+\frac{1}{4e^2_B}- g\right). \label{mW}
\ee

These bosons will be stable if $m^2_Z \geq 0$, therefore taking in account (\ref{la1}), the condition for having stable massive
bosons is that,
\bb
|g| \leq \frac{1}{4} \left( \frac{1}{e^2_A}+\frac{1}{e^2_B}\right). \label{palma}
\ee
Otherwise, the vacuum is not spontaneusly broken and there would not be mass bosons.

The spectrum is then the following: one $U(1)$ massless gauge  field $Z_\mu^3$ (with two polarizations), two massive vector
fields $Z^1_i$ and $Z^2_i$ (three polarizations each) and two massive
vector fields $W^1_i$ and $W^2_i$ (three polarizations each). From this analysis also we get one massless scalar under
rotation boson $\omega$ and  one  massless vector field $W^3_i$. However, If we couple gravity, the last massless vector
fields $W^3_i$ and $\omega$ can be removed by a general coordinate transformation and, therefore, they do not not
contribute  to the spectrum in this sector.

One should note that in the case $g=0$ the gauge symmetry $SU(2)\times SU(2)$ is recovered and, therefore, the mass
terms are forbidden  by the full gauge symmetry even though the mass expression is different from zero. Indeed, the shifting
in the vacuum, (\ref{lifting}), can be removed by a suitable gauge transformation.

\section{Instantons and Kinetic Mixing Terms}
The model discussed above also have classical instanton solutions. Indeed, let us consider --as a
warm-up exercise--   the \lq \lq free" case where the action in terms of two fields  is
\bb
S= \frac{1}{4} \int d^4x\, F^2_{\mu \nu}(A) +  \frac{1}{4} \int d^4x\, F^2_{\mu \nu}(A). \label{insta1}
\ee

Then assuming self-duality conditions $F_{\mu \nu}(A)= {\tilde F}_{\mu \nu}(A)$ and
$F_{\mu \nu}(B)= {\tilde F}_{\mu \nu}(B)$ one finds
\bb
S=8\pi^2 \left(\frac{n_A}{e^2_A} +\frac{n_B}{e^2_B} \right) ,  \label{insta2}
\ee
which is the standard instanton solution for two non-interacting gauge fields.

If we add the  kinetic mixing terms (\ref{inte}) and we use (\ref{fina}) one finds that the relevant part of the action for the
instanton calculations is
\bb
S= \left( \frac{e^2}{8 e^2_A} + \frac{e^2}{8 e^2_B} + \frac{g}{2}\right) \int d^4x F^2_{\mu \nu}(Z). \label{insta3}
\ee

The other terms vanishes in $\mathbb R^4$ when $|x|\rightarrow \infty$ and, therefore, after to use the self-duality condition
for $F_{\mu\nu}(Z)$
one finds
\eqb
S &=&  \left( \frac{1}{8 e^2_A} + \frac{1}{8 e^2_B} + \frac{g}{2}\right) \int d^4x {\tilde F}_{\mu \nu}(Z) F_{\mu \nu}(Z),
\nonumber
\\
&=&\frac{8 \pi^2}{{\tilde e}^2} n. \label{insta4}
\eqf
where the redefined coupling constant is
\bb
{\tilde e}^2 =  \left( \frac{1}{4e^2_A} + \frac{1}{4e^2_B} + 2 g \right)^{-1}. \label{charge}
\ee

Thus, in this $SU(2)$ case,  the coupling constant also is redefined as in the Holdom model. The physical
reasons, however,  are completely different.

\section{final Remarks}

In this paper  a non-abelian extension of the  Holdom model has been proposed.
Contrarily to the abelian counterpart, the term ${\mbox tr} F_{\mu \nu}(A)F_{\mu \nu}(B)$ breaks partially the
$SU(2)\times SU(2)$ symmetry to $SU(2)$ implying interesting new physical properties. Among these new properties
one can point out the following; the partial gauge symmetry breaking implies that only the combination (\ref{con2}) transforms
as a gauge potential. Whereas (\ref{con3}) transforms as a matter field in the adjoint representation.

A careful analysis shows that the model proposed here is compatible with spontaneous symmetry breaking. Therefore,
it provides mass for $W^a_\mu$  as in the standard Goldstone mechanism. Furthermore, the vector character of $W^a_\mu$
induces a spontaneous Lorentz symmetry breaking.

By coupling the $Z^a_\mu$ gauge fields to $W^a_\mu$, some gauge bosons acquire mass as in the Higgs mechanism.
Therefore, the model proposed here becomes equivalent to a vector Higgs mechanism. This fact together with the
spontaneous Lorentz symmetry breaking seems like recent works on  modified gravity \cite{bere} and composite model
discussions \cite{bjorken,philli,atkatz,hosotani}.

Another interesting aspect of this model is that the presence of an instanton solution suggests additional non-trivial properties
of the vacuum. Furthermore, the coupling constat is effectively  redefined ${\tilde e}$.  This last fact could
control possible divergences due to the explicit   symmetry breaking, however, a proof of the renormalizability of the model
proposed here is out of the scope of this paper.

\vspace{0.3 cm}

\noindent\underline{Acknowledgements}:  We would like to thank to F. M\'endez  and F. A. Schaposnik by fruitful
discussions.
This work was partially supported by FONDECYT-Chile grant-1050114 and DICYT (USACH).



\end{document}